# Purcell enhancement of a single silicon carbide color center with coherent spin control

A. L. Crook,[1,2] C. P. Anderson,[1,2] K. C. Miao,[1] A. Bourassa,[1] H. Lee,[1,2] S. L. Bayliss,[1] D. O. Bracher,[3,4] X. Zhang,[3] H. Abe,[5] T. Ohshima,[5] E. L. Hu,[3,4] and D. D. Awschalom[1,2,6, *]

[1]*Pritzker School of Molecular Engineering, University of Chicago, Chicago, IL 60637, USA*
[2]*Department of Physics, University of Chicago, Chicago, IL 60637, USA*
[3]*John A. Paulson School of Engineering and Applied Sciences, Harvard University, Cambridge, MA 02138*
[4]*Department of Physics, Harvard University, Cambridge, MA 02138*
[5]*National Institutes for Quantum and Radiological Science and Technology (QST), 1233 Watanuki, Takasaki, Gunma 370-1292, Japan*
[6]*Center for Molecular Engineering and Materials Science Division, Argonne National Laboratory, Lemont, IL 60439, USA*

Silicon carbide has recently been developed as a platform for optically addressable spin defects. In particular, the neutral divacancy in the 4H polytype displays an optically addressable spin-1 ground state and near-infrared optical emission. Here, we present the Purcell enhancement of a single neutral divacancy coupled to a photonic crystal cavity. We utilize a combination of nanolithographic techniques and a dopant-selective photoelectrochemical etch to produce suspended cavities with quality factors exceeding 5,000. Subsequent coupling to a single divacancy leads to a Purcell factor of ~50, which manifests as increased photoluminescence into the zero-phonon line and a shortened excited-state lifetime. Additionally, we measure coherent control of the divacancy ground state spin inside the cavity nanostructure and demonstrate extended coherence through dynamical decoupling. This spin-cavity system represents an advance towards scalable long-distance entanglement protocols using silicon carbide that require the interference of indistinguishable photons from spatially separated single qubits.

Silicon carbide (SiC) is a technologically mature semiconductor used in commercial applications ranging from high-power electronics to light-emitting diodes. These commercial uses have led to well-developed wafer-scale fabrication processes and precise control of doping during single-crystal growth. Concurrently, SiC has generated interest for low-loss nanophotonics, nonlinear optical phenomena, and micro-electromechanical systems (MEMS)[1–3]. Recently, SiC has also shown promise as a host for optically addressable spin defects. These include the neutral divacancy ($VV^0$), the silicon vacancy ($V_{Si}$), and substitutional transition metal ions ($Cr^{4+}$, $V^{4+}$, $Mo^{5+}$), among others[4–15]. For these defects, isolated electronic states formed in the band gap create spin sublevels. The spin state can then be manipulated with applied microwave fields and read out using their distinct levels of photoluminescence (PL) after optical excitation. Experiments have utilized this optical readout mechanism to demonstrate control of the ground state spin, forming the basis of a qubit[5,8,10,11]. Additionally, the near-infrared emission of many of these SiC defects makes them compatible with existing fiber optic networks that operate at telecom wavelengths. For the $V_{Si}$ and $VV^0$, investigation of the excited state optical fine structure has also revealed spin preserving transitions that can be individually addressed in high-quality samples[16,17]. The combination of ground state spin control with optical spin readout using these transitions lays the foundation for a high-fidelity spin-to-photon interface, with potential applications in quantum communication, distributed quantum computing, and quantum sensing.

For point-defect qubits in semiconductors such as SiC, an overarching goal is the development of a long-distance interconnected quantum network where electron spins act as stationary qubit nodes interconnected by single photons acting as carriers of quantum information[18]. This architecture could then be utilized as a "quantum repeater" to relay quantum states over length scales beyond the ~100 km limit of single photons through fiber[19,20]. However, entanglement rates and scalability are limited by intrinsic emission into the zero-phonon line (ZPL), which is used to produce indistinguishable photons for interference between spatially separated spins. To this end, the defect spin community has explored using photonic nanocavities to enhance a coupled defect's ZPL emission. This enhancement is typically expressed as the Purcell factor, which quantifies an excited state's lifetime reduction as a ratio of emission rates[21]:

$$F \equiv \frac{\Gamma_{cavity}}{\Gamma_{bulk}} = F_1 \cdot F_2 \cdot \frac{3Q}{4\pi^2 V} \left(\frac{\lambda_{cavity}}{n}\right)^3 + 1, \quad (1)$$

Where $\Gamma_{cavity}$ and $\Gamma_{bulk}$ are the cavity-enhanced and unmodified emission rates, with $F = 1$ defining no enhancement. For the photonic cavity, $Q$ is the quality factor, $V$ is the mode volume, $\lambda_{cavity}$ is the resonant wavelength, and $n$ is the index of refraction. The terms $F_1$ and $F_2$ represent spatial overlap and spectral matching between the emitter and cavity mode, respectively, and are both equal to one in the case of perfect coupling (see supplement). In recent work, cavity-defect systems in both diamond and silicon carbide have featured photonic crystal cavities with high quality factors (~$10^3$-$10^4$) and small mode volumes (~$(\lambda/n)^3$) [22–29]. For silicon carbide in particular, its high-Q nanophotonic capabilities[30,31], intrinsic spin defect emitters, and wafer-scale doping control situates it to be a highly promising platform for integrated spin-photonic systems. However, despite SiC's potential, photonic integration with single $VV^0$s remained unexplored.

In this letter, we fabricate nanobeam photonic crystal cavities in 4H-SiC and couple them to single $VV^0$s. We start with a description of the photonic cavity design and fabrication process. We then characterize a single $VV^0$ within the cavity structure at cryogenic temperatures. When the cavity is tuned into resonance with the $VV^0$, we observe a Purcell enhancement of ~50 and an improvement of the Debye-Waller (DW) factor from ~5% to ~70-75%. Lastly, we demonstrate microwave control of the ground-state spin and measure spin coherence times. This union between single defect control and cavity-emitter interactions results in significant increases in the $VV^0$'s ZPL emission with coherent electron spin states, establishing important groundwork for single-shot readout and scalable remote spin entanglement using defect spins.

**Cavity fabrication and characterization.** In order to create a photonic nanocavity, light must be confined in all three dimensions. Archetypal structures employ a sub-micron thin slab of dielectric material to provide out-of-plane confinement through total internal reflection and a patterning of Bragg mirrors to provide in-plane confinement. This results in either a 1D or 2D photonic crystal design, with both systems demonstrating high quality factors with small mode volumes[32,33]. For this work, we select a 1D nanobeam structure due to its more compact size and successful demonstration in previous work[25,26]. We use the general design outlined in work by Bracher *et al.*[26], where circular holes are tapered to ellipses at the center of the nanobeam. This forms a photonic cavity, with a simulation of the resonant mode shown in Fig. 1a and simulated quality factors typically in the range of ~$10^5$-$10^6$.

To form the nanobeam cavities, we utilize electron beam lithography for in-plane patterning and photoelectrochemical (PEC) etching for creating an undercut structure[34,35]. The fabrication procedure, outlined in Fig. 1b, begins with electron beam lithography to define a thin nickel mask with evaporation and liftoff. Next, a $SF_6$-based inductively coupled plasma (ICP) etches through the silicon carbide in the regions not protected by the nickel. After an acid clean to remove the metal, a PEC etch and subsequent HF clean selectively etches the layer of p-type 4H-SiC 400 nm below the top surface, suspending the nanobeams. A scanning electron microscopy (SEM) image of a representative device is shown in Fig. 1c. The nanobeams appear smooth both on the topside and sidewalls of the beams, with relatively smooth and vertical etched holes. We employed a variety of cavity dimensions to create cavity resonances that include ZPLs for each of the (*hh*), (*kk*), and (*kh*) $VV^0$s (see supplement for nomenclature). Several resulting cavities were then characterized for optical resonances with photoluminescence spectra collected using 905 nm excitation. For one such nanobeam, we measured a quality factor of ~5,100 (Fig. 1d) that was typical for photonic cavities in this sample.

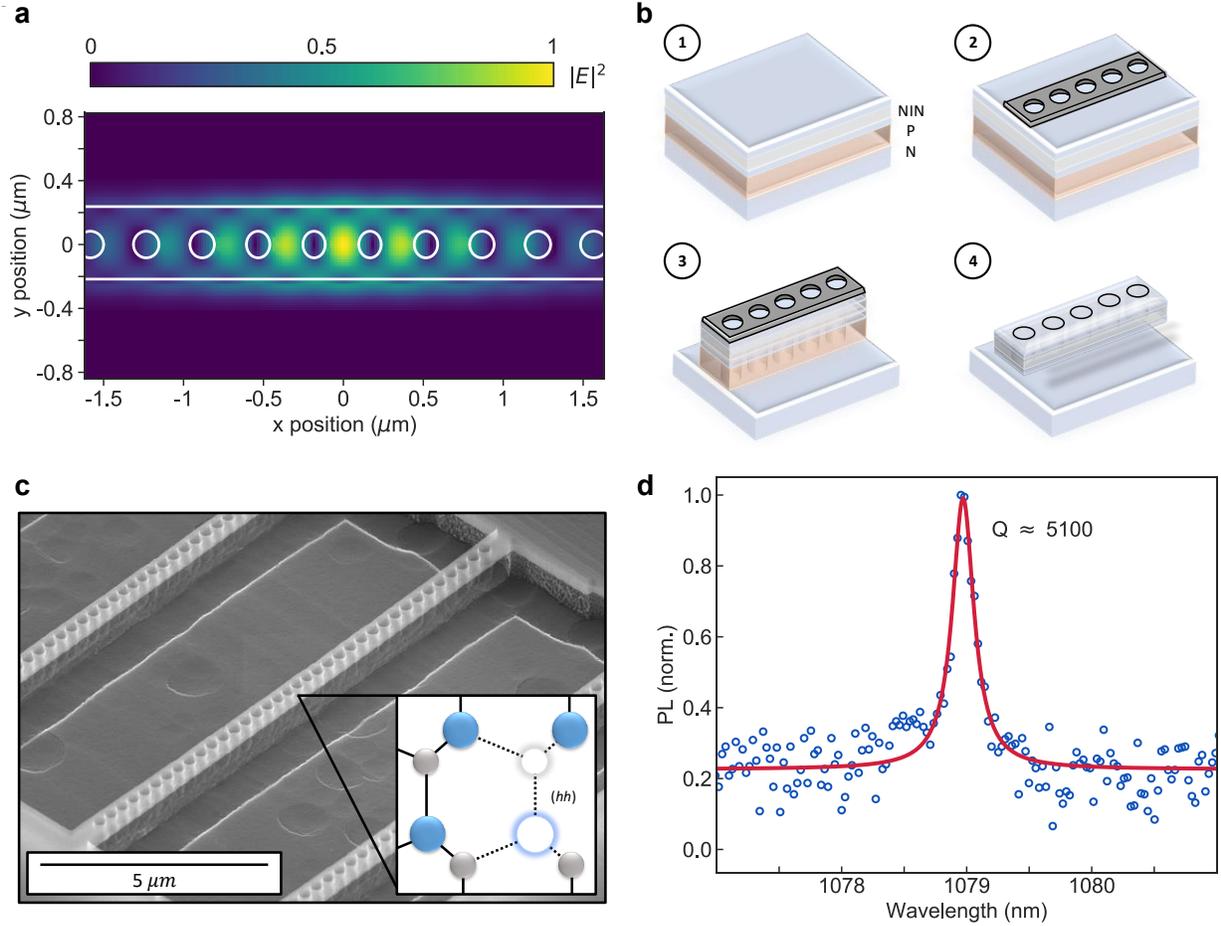

**Figure 1.** Nanobeam photonic cavities in 4H silicon carbide. (a) Simulation of nanobeam cavity mode performed with *Lumerical FDTD*, with a simulated quality factor of Q~3x10$^5$. (b) Outline of fabrication procedure. (1) A NINPN doped SiC chip is used as the starting material. (2) electron beam lithography defines a 25 nm thick nickel mask. (3) An SF$_6$ based inductively coupled plasma (ICP) etch transfers the mask pattern to the SiC substrate. (4) A photoelectrochemical etch (PEC) selectively etches p-type SiC and creates an undercut structure. (c) Scanning electron microscope image of fabricated photonic crystal nanobeam cavities. Inset is a lattice representation of the (*hh*) VV$^0$. (d) Photoluminescence spectrum of a nanobeam cavity taken at room temperature, showing a quality factor of ~5,100 extracted from the full-width half-max of a Lorentzian fit.

**Single VV$^0$ characterization.** After creating defects with an electron irradiation procedure (see supplement), we characterize a single VV$^0$ coupled to the cavity in Fig. 1d. Fig. 2a shows a spatial PL scan taken at 5 K with off-resonant (905 nm) excitation and the cavity off/on resonance with a ~1078 nm VV$^0$ transition. Subsequent photoluminescence excitation (PLE) measurements reveal two peaks at frequencies of 277.984 THz and 278.027 THz (Fig. 2b). We then perform pulsed optically detected magnetic resonance (ODMR) with resonant optical excitation and a nearby wire-bond to drive microwave spin transitions. This results in an ODMR peak centered at 1.328 GHz (Fig. 2c, center), which is closest to the (*hh*) VV$^0$ transition at 1.336 GHz[36]. As we vary the strength of an applied *c*-axis oriented magnetic field, this resonance separates into two lines due to a Zeeman splitting (Fig. 2c). The observed shifts at ~2.76 MHz/G match closely with the electron gyromagnetic ratio of 2.8 MHz/G found in the *c*-axis (*hh*) and (*kk*) defects[36]. The presence of only one ODMR peak under zero magnetic field indicates that the transverse zero-field splitting (*E*) is approximately zero in the VV$^0$ spin Hamiltonian[5], which is also consistent with a *c*-axis oriented VV$^0$. If we instead apply off-resonant optical excitation, we observe ODMR with a negative contrast that matches previous work with (*hh*) VV$^0$s[36] (see supplement). Thus, while the ZPL of this defect matches the (*kh*) VV$^0$ wavelength (~1078 nm), the *c*-axis spin orientation and the off-resonant ODMR contrast sign indicate the presence of an (*hh*) VV$^0$. We attribute this behavior to a highly strained environment (see supplement), likely due to the high doping levels used during growth[37–40].

Additionally, we confirm the presence of a single optical emitter with a second order correlation measurement under resonant excitation (Fig. 2d). The antibunching dip $g^{(2)}(0) \leq 0.5$ indicates the presence of a single emitter, and the value $g^{(2)}(0) = 0.096$ indicates that this VV$^0$ is an excellent single photon source. Meanwhile, the observed bunching behavior is indicative of nonradiative transitions from the excited state. Solving the rate equations for this system (see supplement) and fitting it to the observed $g^{(2)}$ gives an effective dark state lifetime of $\tau_{dark} \approx 75\ ns$. The nonradiative transitions are likely a combination of inter-system crossing (ISC) decays and VV$^0$ ionization. Although the ISC rates have not been explored in 4H-SiC VV$^0$s, in the 3C VV$^0$ they were estimated to be on a similar time scale of ~100 ns[16]. Additionally, VV$^0$ ionization can be observed in our experiment under lower laser powers as a blinking behavior. Without a sufficiently strong 905 nm charge reset pulse, the VV$^0$ may be trapped in a non-radiative charge state for long periods of time, as has been observed in other work[41,42].

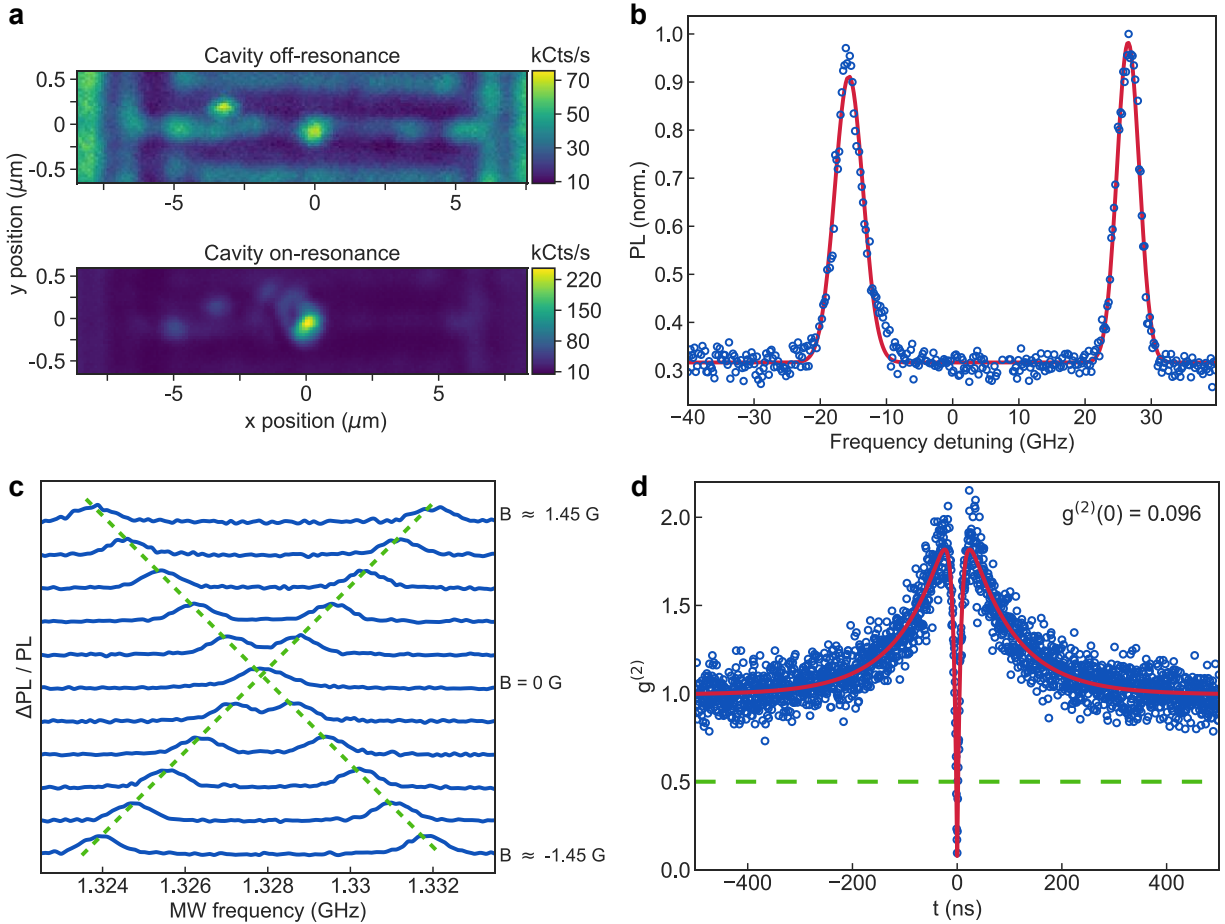

**Figure 2.** Single VV$^0$ spin qubit in a photonic crystal cavity. (a) Spatial photoluminescence scan around a photonic nanobeam cavity under off-resonant excitation with the cavity on (top) and off (bottom) resonance with an embedded VV$^0$. (b) Photoluminescence excitation (PLE) scan over the central VV$^0$ in the nanobeam. Detuning is with respect to 278.000 THz. Peaks are present at 277.984 THz and 278.027 THz with FWHM Gaussian linewidths of 5.02 ± 0.08 GHz and 3.98 ± 0.06 GHz, respectively, with 95% confidence intervals. (c) Optically detected magnetic resonance (ODMR) of the nanobeam VV$^0$ under resonant excitation and varying applied magnetic fields. Separate scans are offset for clarity and the green dotted lines are guides to the eye showing a linear Zeeman splitting. Magnetic field strengths vary between ±1447 mG in uniform steps of ~289 mG. (d) $g^{(2)}(t)$ autocorrelation measurement of the nanobeam $VV^0$, with a best fit (red) including the presence of a nonradiative state and a horizontal line (green) at $g^{(2)} = 0.5$ indicating the upper threshold for a single emitter. The data contains $g^{(2)}(0) = 0.096$ with no background subtraction and the best fit line gives $g^{(2)}(0) = 0.079$. All measurements were taken at 5 K.

**Purcell enhancement.** With a tunable photonic nanocavity and a $VV^0$ emitter within its mode volume, we are able to observe Purcell enhancement of the $VV^0$'s optical emission. When the cavity is off resonance with the $VV^0$ and addressed with an off-resonant 905 nm laser, two peaks at ~1078 nm can be observed in a PL spectrum (Fig. 3a., top inset). These peak locations and their ~40 GHz splitting correspond to the PLE peaks observed under resonant excitation (Fig. 2b). We will label the lower/higher energy transitions as the lower/upper branches of the orbital fine structure, respectively[16]. When the cavity is then tuned into resonance with the defect, a significant increase in emission is observed, with selective enhancement of the lower branch shown in Fig. 3a. This count rate increase correlates closely with the Purcell factor, which in this case is given by:

$$F = \frac{I_{ZPL,on}}{I_{ZPL,off}}, \qquad (2)$$

Where $I_{ZPL,on}$ and $I_{ZPL,off}$ represent the ZPL intensity when the cavity is on resonance and blueshifted off resonance, respectively. This equation matches the form of eq. (1), with ZPL intensities acting as measures of emission rates. Integrating the counts under the two peaks when off and on cavity resonance gives Purcell factors of ~53 (Fig. 3a) and ~16 (see supplement) for the lower and upper branches, respectively. This difference could be explained by differing optical dipole orientations of the two branches, which would give varied matching to the cavity mode. A similar effect was observed for cavity enhancement of $V_{Si}$ defects in silicon carbide, which also displays two rotated optical dipoles[26].

To corroborate the presence of Purcell enhancement, we directly measured excited state lifetimes with the cavity on and off resonance with the $VV^0$. Using resonant excitation pulses from an electro-optic modulator, we observe an off-resonance lifetime of $\tau_{off} = 15.7 \pm 0.3\ ns$ (consistent with bulk measurements[16]) and an on-resonance lifetime of $\tau_{on} = 5.3 \pm 0.1\ ns$ (Fig. 3b). The relationship between measurable lifetimes and the Purcell factor is given by:

$$F = \frac{\tau_{dark}(\tau_{off} - \tau_{on})}{\alpha \tau_{on}(\tau_{dark} - \tau_{off})} + 1, \qquad (3)$$

Where, for the $VV^0$, $\tau_{dark}$ is the combined lifetime of all nonradiative decays, $\tau_{off}$ is the lifetime off cavity-resonance, $\tau_{on}$ is the lifetime on cavity-resonance, and $\alpha$ is the intrinsic DW factor (see supplement). Combining our measurements with a previously measured ~5.3% DW factor[16] gives a Purcell factor of $F \approx 48$, which is in good agreement with the value of $F \approx 53$ from spectral measurements.

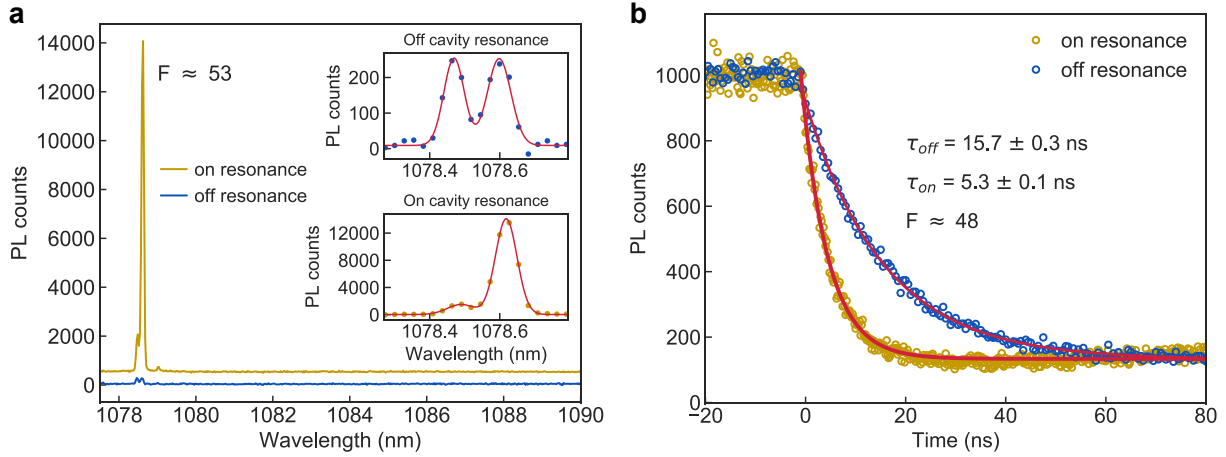

**Figure 3.** Purcell enhancement of a single VV$^0$ in a photonic crystal cavity. (a) Emission spectrum of the VV$^0$ when excited with off-resonant 905 nm laser light with the cavity on (inset, lower right) and off (inset, upper right) resonance with the lower energy branch. A ratio of emission intensities gives a Purcell factor of ~53. The on-resonance trace for the combined plot is vertically offset for clarity. (b) Lifetime measurements of the VV$^0$ under resonant 277.984 THz excitation with the cavity on and off resonance. Fits to an exponential decay of $\exp(-t/\tau)$ give a shortened lifetime ($\tau = 5.3 \pm 0.1\ ns$) when on resonance vs. off resonance ($\tau = 15.7 \pm 0.3\ ns$) with 95% confidence intervals, giving a Purcell factor of ~48. All measurements were taken at 5 K.

One of the direct consequences of a Purcell enhancement is an increased Debye-Waller factor, which follows the relation:

$$F = \frac{\beta(\alpha - 1)}{\alpha(\beta - 1)}, \qquad (4)$$

Where $\alpha$ and $\beta$ represent the VV$^0$'s DW factor off and on cavity resonance, respectively (see supplement). For our sample, spatially varying background luminescence from NV centers in the n-doped silicon carbide[6,14,15,43] makes it difficult to directly integrate spectrometer counts to obtain $\alpha$ and $\beta$. However, we do observe background subtracted count rates of 120 kCts/s and 460 kCts/s when off and on cavity resonance, which allows us to estimate an on-resonance DW factor of $\beta \approx 75\%$ and a Purcell factor of $F \approx 54$ (see supplement). These numbers match well with the 74% factor obtained from equation (5), and the Purcell factors of ~53 and ~48 obtained from Fig 3. Given the agreement between these independent measurements, we infer that the VV$^0$ emits 70-75% into the ZPL when the cavity is on resonance. This is a significant improvement over the intrinsic ~5% DW factor for the VV$^0$, and the Purcell-enhanced 460 kCts/s is among the highest count rates achieved for SiC spin defects. Combined with the lifetime reduction, this enhancement greatly aids in achieving single-shot readout and high entanglement rates in SiC.

**Coherent spin control.** While addressing the cavity VV$^0$ with resonant microwave pulses and resonant optical excitation, we drive coherent Rabi oscillations between the spin sublevels. To address a single microwave transition, we apply a small magnetic field of ~6 G parallel to the *c*-axis to Zeeman split the spin resonances and then focus on the $|0\rangle$ to $|+1\rangle$ transition at 5 K. Under these conditions, we observe Rabi oscillations with a readout contrast of ~40% (Fig. 4a). This contrast level is significantly higher than the typical 10-15% observed for off-resonant Rabi oscillations[5], but below the ~94-98% levels observed with resonant excitation[16,41,44]. This indicates that individual optical spin transitions are moderately selective, but still display a spectral overlap from the ~4-5 GHz PLE optical linewidths broadened from spectral diffusion. Given that individual spin transitions for c-axis VV$^0$s are typically separated by a few GHz[16], it should be possible to achieve higher contrast with a slight narrowing of linewidths.

We then apply Ramsey interferometry and Hahn echo pulse sequences on the same $|0\rangle \to |+1\rangle$ transition to determine the spin dephasing and spin coherence times. Under a c-axis magnetic field of ~218 G, we obtain a dephasing time of $T_2^* = 592 \pm 18 \; ns$ (Fig. 4b) and a decoherence time of $T_2 = 9.3 \pm 2.0 \; \mu s$ (Fig. 4c). Under a lower magnetic field of ~6 G, we obtain similar times of $T_2^* = 605 \pm 33 \; ns$ and $T_2 = 7.4 \pm 0.6 \; \mu s$ (see supplement), indicating that coherence in this sample is not primarily limited by the SiC nuclear spin bath[45]. Collectively these times are shorter than previous reports of $T_2^* \approx 1\text{-}2 \; \mu s$ and $T_2 \approx 1.2 \; ms$ in bulk SiC c-axis VV$^0$s [5], with the discrepancy likely arising from magnetic dipole interactions with electron spins from n-type dopants and surface charge traps. It is worth noting that for a VV$^0$ located in the NIN epilayer without fabricated structures, we measure $T_2^* = 4.01 \pm 0.38 \; \mu s$ and $T_2 = 200 \pm 27.6 \; \mu s$ under ~218 G (see supplement). Therefore, it appears that a combination of nearby doping and fabricated surfaces results in shortened coherence times. However, there is a variety of approaches to offset these effects. The PEC undercut could likely be carried out at lower doping levels, and post-fabrication surface treatments could potentially be used to limit the presence of surface charge traps.

In the regime where $T_1$ is greater than $T_2$, it should be possible to extend spin coherence through dynamical decoupling sequences. For the cavity VV$^0$, we observe no appreciable spin relaxation over a 100 μs time scale, placing a lower bound of $T_1 > 100 \; \mu s$ and indicating that $T_2$ is not $T_1$ limited. This is to be expected for the VV$^0$ at cryogenic temperatures, where $T_1$ has been measured to be at least 8-24 ms at 20 K[36]. As a proof of principle, we then employ a Carr-Purcell-Meiboom-Gill (CPMG) sequence[46] at low field with one, two, and four π pulses (Fig. 4d). Stretched exponential fits give $T_2$ values of $6.8 \pm 0.7 \; \mu s$, $11.0 \pm 1.9 \; \mu s$, and $19.5 \pm 6.1 \; \mu s$, indicating the viability to extend spin coherence in SiC nanostructures with dynamical decoupling.

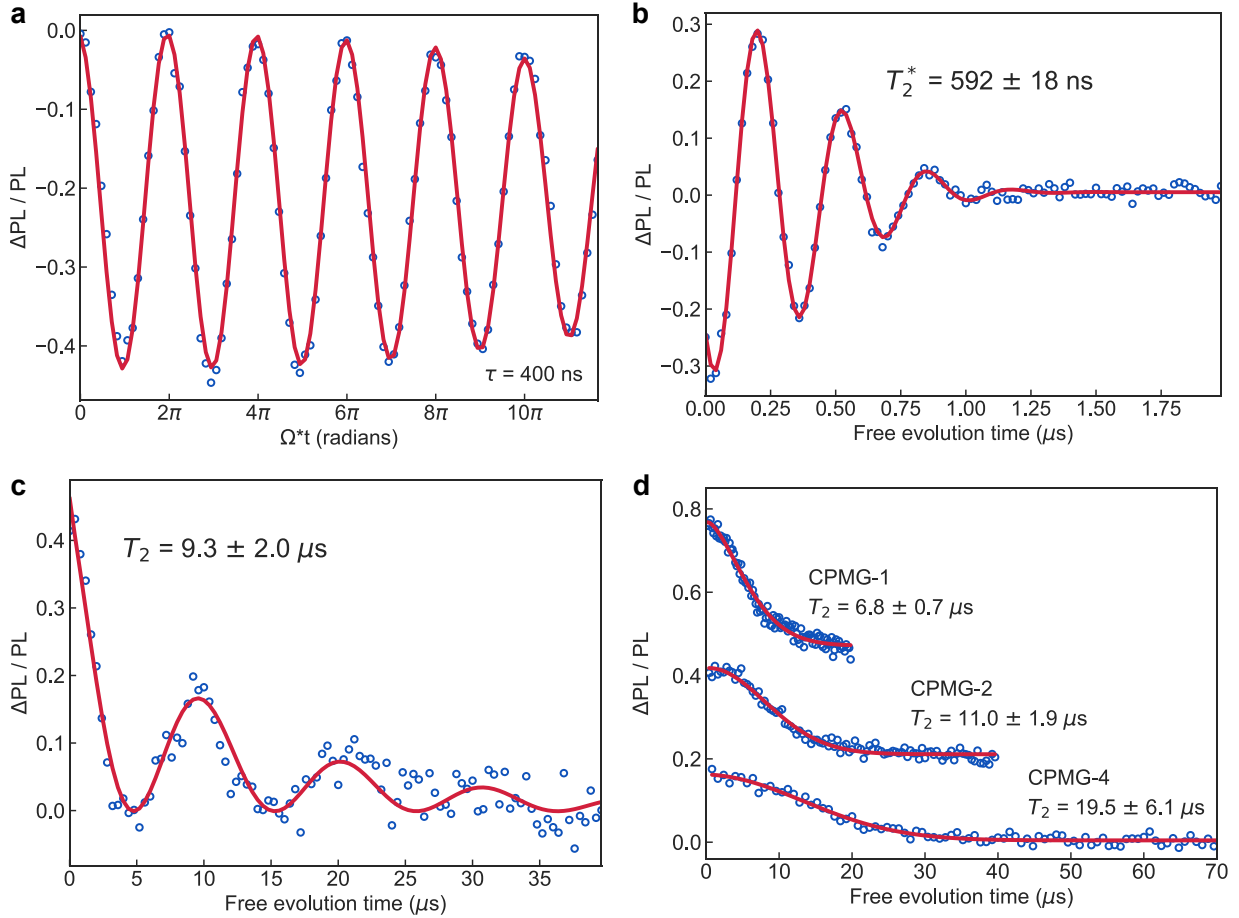

**Figure 4.** Coherent spin control of VV$^0$ in photonic crystal cavity. (a) Rabi oscillations between the $|0\rangle$ and $|+1\rangle$ states under a ~6 G $c$-axis magnetic field. The MW pulse length is kept constant at 400 ns while the applied power is increased. (b) Ramsey sequence collected 3 MHz detuned from resonance at a ~218 G $c$-axis magnetic field. Points are fitted to a sinusoid decaying as $\exp(-(t/T_2^*)^n)$, giving $T_2^* = 592 \pm 18$ $ns$ and $n = 2.08 \pm 0.19$ with 95% confidence intervals. (c) Hahn echo sequence collected under a ~218 G $c$-axis magnetic field. Points are fitted to a $\sin^2$ decaying as $\exp(-(t/T_2)^n)$, giving $T_2 = 9.3 \pm 2.0$ $\mu s$ and $n = 0.80 \pm 0.17$ with 95% confidence intervals. (d) Dynamical decoupling with Carr-Purcell-Meiboom-Gill (CPMG) sequences to extend the T$_2$ decay time. Taken under a ~6 G $c$-axis magnetic field. Fits to $A\exp(-(t/T_2)^n)$ give $T_2 = 6.8 \pm 0.7$ $\mu s$, $T_2 = 11.0 \pm 1.9$ $\mu s$, and $T_2 = 19.5 \pm 6.1$ $\mu s$ for CPMG-1 (a regular Hahn echo sequence), CPMG-2, and CPMG-4, respectively. Corresponding $n$ values are $n = 1.6 \pm 0.2$, $n = 2.0 \pm 0.4$, and $n = 2.1 \pm 0.6$. Confidence intervals are one standard deviation for the CPMG data. All data were collected on the $m_s = |0\rangle \rightarrow |+1\rangle$ transition. All measurements were taken at 5 K.

**Discussion.** Experimentally, increases in both the Debye-Waller factor and PL count rate have significant implications for enhancing the entanglement generation rate between VV$^0$ spins. In the Barrett-Kok protocol[47], for example, the entanglement success rate is proportional to the square of the DW factor for two consecutive ZPL detection events. Using the ~75% DW factor measured here then gives a significant projected entanglement rate increase of $(0.75/0.053)^2 \approx 200$ between two ($hh$) VV$^0$s. Additionally, entanglement verification relies on single-shot readout to determine the electron spin state in a single measurement[48], which is ultimately limited by photon detection throughput. For this system, the Purcell-enhanced threefold lifetime decrease would correspond to triple the emission events before a spin flip. The overall increase of off-cavity-resonance PL (~120 kCts/s) compared to bulk VV$^0$s (typically 40-50 kCts/s) is also indicative of a slightly improved collection efficiency. This is a vital factor for single-shot readout measurements, since a majority of PL from bulk VV$^0$s is lost due to reflection and refraction at the

SiC/air interface. Thus, both single-shot readout of $VV^0$ spin states in a photonic structure and photonically enhanced entanglement could be achievable in future studies.

The cavity-enhanced $VV^0$ studied here contains 4-5 GHz optical linewidths comparable to those seen in near surface NV centers in diamond[49], but above the lifetime limit of ~11 MHz[16]. We attribute the broadened optical linewidths to spectral diffusion originating from a fluctuating charge environment around the defect. These charge fluctuations could be from nearby doped regions, other nearby defects and impurities, or surface charge traps. It is worth noting that the optical linewidths are broader in 400 nm suspended I-type SiC (~10-20 GHz) compared to the suspended NIN shown here (~4-5 GHz). Thus, doping configurations and growth conditions can have significant effects on spectral diffusion, opening the possibility to achieve narrow linewidths through properly doped structures. Additionally, optical linewidths are ~1 GHz for defects in the intrinsic layer of NINPN material before fabrication (see supplement), indicating that the fabrication process or final nanostructure is a significant source of broadening. To counteract this effect, surface treatments[50] or applied voltages could be used to maintain narrow linewidths. Under applied electric fields, for example, $VV^0$ optical linewidths as narrow as ~20 MHz have been observed[41,44].

In conclusion, we have fabricated a photonic crystal cavity in silicon carbide coupled to a single $VV^0$. We observe Purcell enhancement of the ZPL optical transition with a Purcell factor of ~50, a subsequent increase in Debye-Waller factor from ~5% to ~70-75%, and coherent spin control of the $VV^0$ ground state with coherence extension. The use of a doped nanostructure allows for potential electric field and charge control, *in situ* Stark tuning, and improved collection efficiencies for optimized geometries, all of which would provide further improvements to the $VV^0$ optical properties. As a whole, this system advances the robustness of spin-to-photon transduction for the $VV^0$ in a technologically mature material. Looking beyond to many-qubit architectures, photonic nanocavities will be a necessary component to maintain scalability across long-distance entanglement networks.

## Associated content
**Supporting information**
Supplementary information accompanies this paper.

## Author information

**Corresponding Authors**
*E-mail: awsch@uchicago.edu

**Author contributions**
A.C. developed, fabricated, and measured the photonic crystal devices. C.A. and H. L. aided with fabrication procedures. C.A., K.M., and A.B. aided with optical characterization, cryogenic spin measurements, and analysis of data, S.B. assisted with resonant lifetime measurements and development of the three-level $g^{(2)}$ model. D.B., X.Z., and E.H. were instrumental in the development of both the PEC etch of SiC and SiC photonic devices. H.A. and T.O. performed electron irradiation of SiC samples to create divacancies. D.A. oversaw and directed the project. A.C., C.A., K.M., A.B., H.L., and S.B. all contributed with the drafting of the manuscript.

**Notes**
The authors declare no competing interests.

## Acknowledgements
This work was supported by the NSF EFRI AQUIRE EFMA-1641099 and the U Chicago MRSEC DMR-1420709. A.C. This work was funded by the U.S. Department of Energy, Office of Science, Office of Basic Energy Sciences. This work also made use of the Pritzker Nanofabrication Facility part of the Pritzker School of Molecular Engineering at the University of Chicago, which receives support from Soft and Hybrid Nanotechnology Experimental (SHyNE) Resource (NSF ECCS-1542205), a node of the National Science Foundation's National Nanotechnology Coordinated Infrastructure. Electron irradiation was funded by the grants JSPS KAKENHI 17H01056 and 18H03770. This work was completed in part with resources provided by the University of Chicago's Research Computing Center.